\documentclass[12pt]{article}

\usepackage{amsmath}
\usepackage{times}
\usepackage{graphicx}
\usepackage{color}
\usepackage{multirow}
\usepackage[authoryear,round]{natbib}
\usepackage{rotating}
\usepackage{latexsym}

\newcommand{\captionfonts}{\normalsize}

\makeatletter  
\long\def\@makecaption#1#2{%
  \vskip\abovecaptionskip
  \sbox\@tempboxa{{\captionfonts #1: #2}}%
  \ifdim \wd\@tempboxa >\hsize
    {\captionfonts #1: #2\par}
  \else
    \hbox to\hsize{\hfil\box\@tempboxa\hfil}%
  \fi
  \vskip\belowcaptionskip}
\makeatother   

\newtheorem{pro}{Proposition}[section]
\newcommand{\argmin}{\mathop{\rm argmin}\limits}	
\newcommand{\finish}{\hfill$\Box$\par}

\begin{document}
\hspace{13.9cm}1

\ \vspace{20mm}\\

\begin{center}
{\LARGE Spontaneous Clustering via Minimum $ \gamma $-divergence}
\end{center}

\ \\
{\bf \large Akifumi Notsu}\\
{notsu@ism.ac.jp\\Department of Statistical Science, The Graduate University for Advanced Studies, \\Tachikawa, Tokyo 190-8562, Japan}\\

\noindent
{\bf \large Osamu Komori}\\
{komori@ism.ac.jp\\The Institute of Statistical Mathematics, Tachikawa, Tokyo 190-8562, Japan}\\

\noindent
{\bf \large Shinto Eguchi}\\
{eguchi@ism.ac.jp\\The Institute of Statistical Mathematics and The Graduate University for Advanced Studies, Tachikawa, Tokyo 190-8562, Japan}\\
%

{\bf Keywords:} cluster analysis, clustering, divergence, $\gamma$-divergence, power entropy

\thispagestyle{empty}
\markboth{}{NC instructions}
\ \vspace{-0mm}\\
%

\begin{center}
{\bf Abstract}
\end{center}
We propose a new method for clustering based on the local minimization of the $ \gamma $-divergence, which we call the spontaneous clustering.  
The greatest advantage of the proposed method is that it automatically detects the number of clusters that adequately reflect the data structure. In contrast, exiting methods such as $K$-means, fuzzy $c$-means, and model based clustering need to prescribe the number of clusters. 
We detect all the local minimum points of the $ \gamma $-divergence, which are defined as the centers of clusters. 
A necessary and sufficient condition for the $ \gamma $-divergence to have the local minimum points is also derived in a simple setting. 
A simulation study and a real data analysis are performed to compare our proposal with existing methods.  

\section{Introduction}

Cluster analysis is a common procedure for grouping similar objects in unsupervised learning \citep{label3134,label2910,label9944}. 
The procedure stably produces a classification, and is frequently used as a preprocessing before supervised learning. 
Cluster analysis has wide applications over many disciplines in exploratory data analysis. 
See, for example, \citet{label6261} and \citet{label5766} for recent developments. There are mainly two approaches in cluster analysis. One is the hierarchical approach which describes a tree structure called dendrogram. The other is the approach of data space partition such as $K$-means algorithm. This paper focuses on the latter approach from a view point of statistical pattern recognition.

We propose what we call the spontaneous clustering. It starts with finding centers of clusters in a data set. For this purpose, we employ a loss function derived from the power entropy with the power index $ \gamma $. It is referred to the $ \gamma $-loss function \citep{label6557,label3582}. Here is a motivational example for the proposal of the spontaneous clustering. Consider the problem of estimating Gaussian mean parameter $ \mu $. The maximum likelihood estimator (MLE) of $ \mu $ is given by the arithmetic mean of the data set as the unique maximum point of the log likelihood function. It is known that the MLE poorly behaves in various situations where Gaussianity assumption {is inappropriate. 
For example, the log likelihood function suggests rather a misleading summary as seen in panel (a) of Figure 1. Alternatively, the $ \gamma $-loss function properly reflects the data shape. For the same data set in panel (a) of Figure 1, panel (b) shows that the $ \gamma $-loss function has two local minimum points corresponding to the two normal distributions. We will propose to determine the centers of clusters by such local minimum points.

Almost all procedures via data space partition need the number of clusters a priori. The selection of the number of clusters is a major challenge in cluster analysis. 
A lot of methods have been proposed in the literature \citep{label2910}. Our clustering method can find the number of clusters automatically as long as the value of $ \gamma $ is properly fixed. The name of the spontaneous clustering comes from this property. Instead of the number of clusters, the value of power index $ \gamma $ should be determined. We will propose two methods to accomplish this aim. One is a heuristic choice of $ \gamma $ that merely relies on the range of the data, and the other is a more sophisticated method based on Akaike Information Criterion (AIC).

This paper is organized as follows. Section 2 describes the algorithm of the spontaneous clustering and selection procedure of the value of $ \gamma $. In section 3 the existence of the local minimum points is discussed. Section 4 investigates the numerical properties of the spontaneous clustering. In section 5 a real data analysis is given. Further a discussion is presented in section 6. 

\begin{figure}[e]
\begin{center}
\includegraphics[width=140mm]{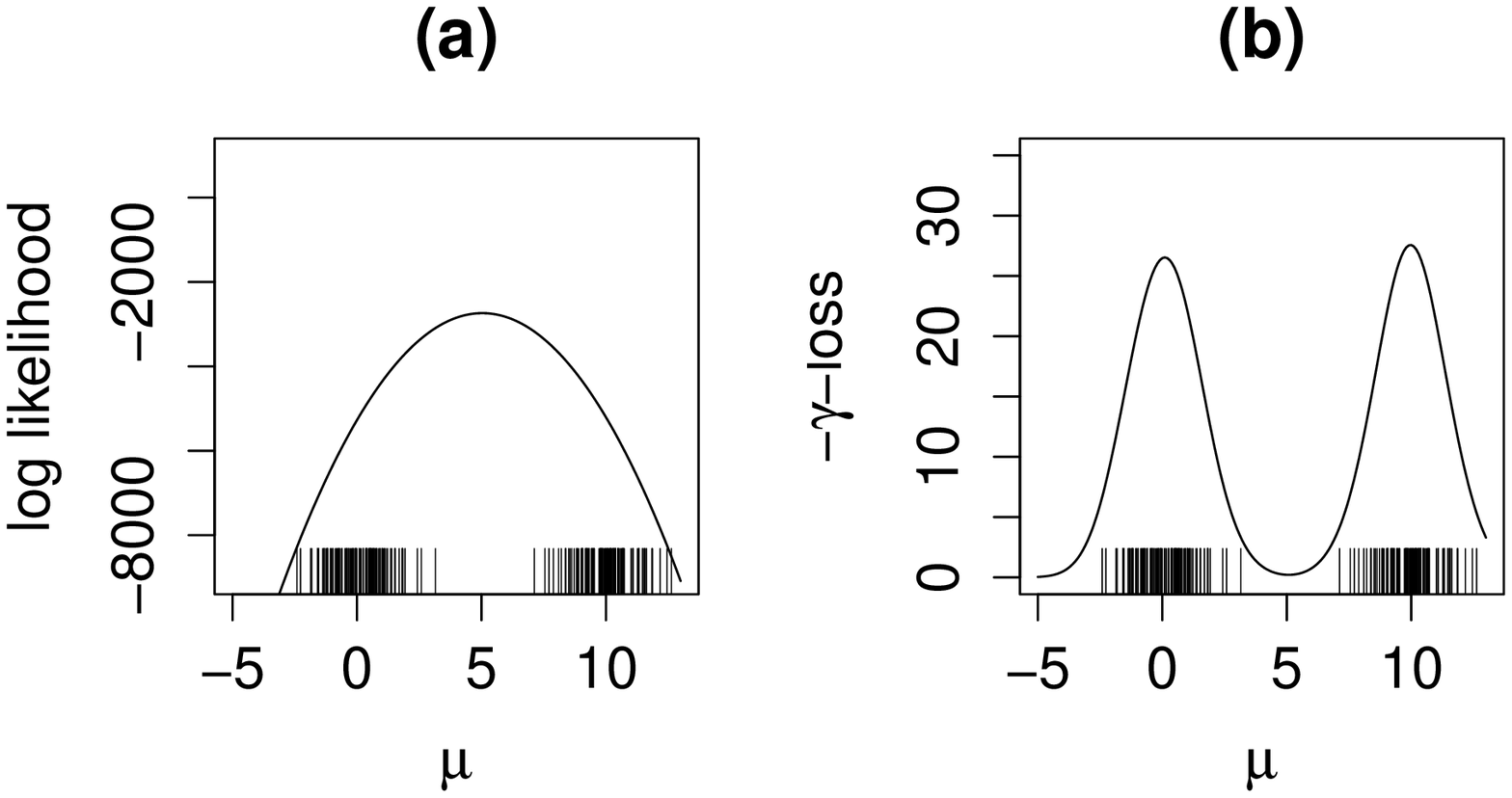}
\end{center}
\caption{(a) Log likelihood function. (b) Minus $\gamma $-loss function ($ \gamma = 1 $). In panels (a) and (b) the data of size 200 is generated from the mixture of two standard normal distributions centered at 0 and 10, respectively.} \label{robust and clustering} 
\end{figure}

\section{Spontaneous Clustering}\label{Spontaneous Clustering}
We begin with a statistical formulation of cluster analysis. Suppose the $ p $-dimensional density function of the population distribution is given by  
\begin{eqnarray}\label{mixture}
g ( x ) = \sum _ { k = 1 } ^ K \tau _ k f _ k ( x ), \ \sum _ { k = 1 } ^ K \tau _ k = 1, \ \tau _ k  > 0 , \ k = 1 , \ldots , K,
\end{eqnarray}
where $ f _ k ( x ) $ is a density function. Let $ \{ x _ 1 , \ldots , x _ n \} $ be a data set generated from $ g $. We apply the $ \gamma $-estimation method to this data set. The $ \gamma $-loss function for the normal distribution with the identity covariance matrix is given by 
\begin{eqnarray}\label{gamma-loss}
L _ \gamma ( \mu ) = - \frac{ 1 }{ n } \sum _ { i = 1 } ^ n \exp \left( - \frac{ \gamma }{ 2 } \| x _ i - \mu \| ^ 2 \right) , 
\end{eqnarray}
apart from a constant, where $ \mu $ and $\| \cdot \| $ denote the mean vector and the Euclidean norm, respectively. In the remainder of the paper, we omit a constant term that does not affect the optimization. 
In panel (b) of Figure \ref{robust and clustering}, $ L _ \gamma ( \mu  ) $ is illustrated. See appendix B for a general introduction to the $ \gamma $-loss function. It is expected that the $ \gamma $-loss function $ L _ \gamma ( \mu ) $ has $ K $ local minimum points corresponding to $ K $ mean vectors with respect to $ f _ 1, \ldots , f _ K $. Then we expect that the local minimum points can help us to define the centers of $ K $ clusters and to build $ K $ clusters in a similar way to the $K$-means algorithm. 
The covariance structure of the data set is taken into consideration in a subsequent discussion.

\subsection{$\gamma$-loss Function for the Normal Distribution}\label{gamma-loss-normal}
We consider the $ \gamma $-loss function for the normal distribution with mean vector $ \mu $ and covariance matrix $ \Sigma $,
\begin{eqnarray*}
L _ \gamma ( \mu , \Sigma ) = - \det \Sigma ^ { - \frac{ \gamma }{ 2 ( 1 + \gamma ) } } \sum _ { i = 1 } ^ n \exp \left( - \frac{ \gamma }{ 2 } ( x _ i - \mu ) ^ \top \Sigma ^  { -1 } ( x _ i - \mu ) \right).
\end{eqnarray*} 
An iteration algorithm to find the local minimum points of $ L _ \gamma ( \mu , \Sigma ) $ is proposed in \cite{label6557} and \citet{label3582}. It is obtained by differentiating $ L _ \gamma ( \mu , \Sigma ) $ with respect to $ \mu $ and $ \Sigma ^ { -1 } $ and setting the derivatives to $ 0 $. The algorithm is a concave-convex procedure (CCCP) \citep{label2602}, so that it is guaranteed to decrease the $ \gamma $-loss function monotonically as the iteration step $t$ increases. It is described as follows. 
\begin{description}
\item[Step 1] 	Set appropriate $ \mu _ 0 $ and $ \Sigma _ 0 $ as initial values.
\item[Step 2] 	Given $ \mu _ t $ and $ \Sigma _ t $, calculate $ \mu _ { t + 1 } $ and $ \Sigma _ { t + 1 } $ by the following update formula,
\begin{eqnarray}
\mu _ { t + 1 } &=& \sum _ { i = 1 } ^ n w _ \gamma ( x _ i , \mu _ { t } , \Sigma _ t ) x _ i , \label{mean-equation} \\
\Sigma _ { t + 1 } &=& ( 1 + \gamma ) \sum _ { i = 1 } ^ n w _ \gamma ( x _ i , \mu _ t , \Sigma _ t ) ( x _ i - \mu _ {t+1} ) ( x _ i - \mu _ {t+1} ) ^ \top , \label{covariance-equation}
\end{eqnarray}
where 
\begin{eqnarray*}
w _ \gamma ( x , \mu , \Sigma ) = \frac{ \exp \left( - \frac{ \gamma }{ 2 } ( x - \mu ) ^ \top \Sigma ^ { -1 } ( x - \mu ) \right) } { \sum _ { j = 1 } ^ n \exp \left( - \frac{ \gamma }{ 2 } ( x _ j - \mu ) ^ \top \Sigma ^ { -1 } ( x _ j - \mu ) \right) } .
\end{eqnarray*}

\item[Step 3] 	For a sufficiently small number $\varepsilon $, repeat Step 2 while 
$$
\| \mu _ { t + 1 } - \mu _ t \| + \| \Sigma _ { t + 1 } - \Sigma _ t \| _ \text{F} < \varepsilon ,
$$ 
where $ \| \cdot \| _ \text{F} $ denotes the Frobenius norm. 
\end{description}
If $ \gamma = 0 $, then the right hand sides of equations (\ref{mean-equation}) and (\ref{covariance-equation}) are equal to the sample mean vector and covariance matrix, respectively, which are nothing but the MLEs. If our aim is to obtain the local minimum points of $ L _ \gamma ( \mu  ) $, then we only have to update $ \mu _ t $ and fix $ \Sigma _ t$ to be the identity matrix $I$. Similarly if our aim is to obtain the local minimum points of $ L _ \gamma ( \mu , \Sigma ) $ with fixed $ \mu $, then we only have to update $ \Sigma _ t $ and fix $ \mu _ t = \mu $.

\subsection{Algorithm of the Spontaneous Clustering}
In general, the spontaneous clustering based on a density function $ f ( x , \theta ) $ with parameter $ \theta $ is defined as follows. 
\\[0.3cm]
\noindent
{\bf Spontaneous Clustering}
\begin{description}
\item[Step 1] 	Find the local minimum points of $ L _ \gamma ( \theta ) $, denoted by $ \hat{\theta} _ 1, \ldots , \hat{\theta}_K $, where $ L _ \gamma ( \theta ) $ is the $ \gamma $-loss function for $ f ( x , \theta ) $.
\item[Step 2] Consider $ K $ clusters according to $ \hat{\theta} _ 1, \ldots , \hat{\theta}_K $, and assign the data to the clusters.
\end{description}
In a special case, the spontaneous clustering based on the normal distribution is defined as follows. We set $ \Theta _ \mu $ and $ \Theta _ { (\mu , \Sigma) } $ are the empty sets at the start of the algorithm. The algorithm of subsection 2.1 is employed in the spontaneous clustering below. 
\\[0.3cm]
\noindent
{\bf Spontaneous Clustering Based on the Normal Distribution}
\begin{description}
\item[Step 1-1] If $ \Theta _ \mu $ is the empty set, choose $ M $ initial values $ x _ { ( 1 ) } , \ldots , x _ { ( M ) } $ in the data set $\{  x _ 1 , \ldots , x _ n \} $ at random. Otherwise, choose initial values in $\{  x _ 1 ,\ldots , x _ n \} $ as follows: $ x _ { ( 1 ) } , \ldots , x _ { ( M ) } $ are $ M $ maximum points of $ d ( \cdot , \Theta _ \mu ) $, where  
$$
d( x , \Theta _ \mu ) = \min _ { \hat{ \mu }  \in \Theta _ \mu } \|x - \hat{ \mu } \| .
$$
\item[Step 1-2] Apply the algorithm in subsection \ref{gamma-loss-normal} to the data set $M$ times with each initial value $x _ { ( i ) } , i = 1 , \ldots, M $ to find the local minimum points of $ L _ \gamma ( \mu ) $. Then add the obtained local minimum points to $ \Theta _ \mu $. 
\item[Step 1-3] Repeat Step 1-1 and 1-2 until the number of elements in $ \Theta _ \mu $ does not increase.
\item[Step 1-4] For each local minimum point $ \hat{ \mu } \in \Theta _ \mu $, obtain a minimum point of $ L _ \gamma ( \hat{ \mu } , \Sigma ) $ with respect to $ \Sigma $, denoted by $ \hat{ \Sigma } $, with the algorithm in subsection \ref{gamma-loss-normal}. Then add $ ( \hat{ \mu } , \hat{ \Sigma } ) $ to $ \Theta _ { ( \mu , \Sigma ) } $.
\item[Step 2] Write $ \Theta _ { ( \mu , \Sigma ) } $ by $\{ ( \hat{ \mu } _ k , \hat{ \Sigma } _ k ) \} _ { k = 1 } ^ K  $ and assign each observation $ x _ i $ to the $ \hat{k} $-th cluster with 
$$
\hat{k}= \argmin _ { k=1,\ldots,K } ( x _ i - \hat{ \mu } _ k ) ^ \top \hat{ \Sigma } _ k  ^ { -1 } ( x _ i - \hat{ \mu } _ k ) . 
$$
\end{description}
In the algorithm of the spontaneous clustering, we define $ ( \hat{ \mu } _ k , \hat{ \Sigma } _ k ) , k = 1 , \ldots , K  $ as the centers and the covariance matrices of clusters. In the remainder of this paper, we focus on the spontaneous clustering based on the normal distribution.

\subsection{Selection Procedure for $ \gamma $} \label{select-gamma}
The value of power index $ \gamma $ plays a key role in the spontaneous clustering, because $ \gamma $ affects the number of clusters obtained by the spontaneous clustering. 
We propose two methods to select the value of $ \gamma $. 
One is a heuristic choice of $ \gamma $ that depends on the range of the data. Our proposal is $ \hat{ \gamma } = 72 / R ^ 2 $, where $ R $ is defined by the maximum range: 
$$
R  = \max _ { j = 1 , \ldots , p } \left\{ \left( \max _ { i = 1 , \ldots , n } x _ { i j } \right) - \left( \min _ { i = 1 , \ldots , n } x _ { i j } \right)  \right\},
$$
where $ x _ i = ( x _ { i 1 } , \ldots , x _ { i p } ) ^ \top $. The outline of the derivation of $ \hat{ \gamma } $ is as follows. 
Suppose the data set is generated from the mixture of two normal distributions centered at $ \mu _ 1 $ and $ \mu _ 2 $ with the identity covariance matrix and the same mixing proportion, respectively. 
Our simulation result suggests that if $ \| ( \mu _ 1 - \mu _ 2 ) / 2 \| = 3\sqrt{2}/2\doteq 2.12 $, then the value of $ \gamma $ needs to be more than or equal to 1 for two local minimum points of $ L _ \gamma ( \mu ) $ to exist. Proposition \ref{exist-two-local} tells that if all the data are multiplied by a scalar $ a $ and the spontaneous clustering is applied to the transformed data, then the value of $ \gamma $ needs to be more than or equal to $ a^{-2} $ to guarantee the existence of two local minimum points of $ L _ \gamma ( \mu ) $. If $ \| ( \mu _ 1 - \mu _ 2 ) / 2 \| = r $, then $ a = r / ( 3 \sqrt{2} / 2 ) $. Hence we propose to use the value of $ \gamma $ defined as 
\begin{eqnarray} \label{estimate-gamma}
\hat{\gamma} = \left( \frac{r} { \frac{ 3 \sqrt{2} }{ 2 } } \right) ^ { - 2 } = \frac{ 9 }{ 2 r ^ 2 } .
\end{eqnarray}
The value of $ r $ can be estimated by the range of the data. Let $ R _ j $ be the range of the $ j $-th variable. If there are $ K $ disjoint clusters lying side by side on a line parallel to the axis of the $ j $-th variable, then we can estimate $ r $ by $ R _ { j } / ( 2 K ) $ as is just illustrated in Figure \ref{two-clusters}. 
There are $p$ variables, so $p$ directions have to be considered simultaneously. We use the maximum range $ R $, and estimate $ r $ by $ R / ( 2 K ) $.  
The value of $ K $ can be determined from our prior knowledge about the possible number of clusters. If $ K = 2 $, we have $ \hat{ \gamma } = 72/R^2 $.
We observe that this rule works well in several empirical studies although the discussion does not completely have the theoretical background.

We also propose a more sophisticated method based on AIC. The value of $ \gamma $ which minimizes $\text{AIC}$ is recommended as the optimal selection of $ \gamma $. 
Let $ K _ \gamma $ be the number of clusters and $ ( \hat{ \mu } _ { \gamma k } , \hat{ \Sigma } _ { \gamma k } ) , k = 1, \ldots , K _ \gamma $ be the centers and the covariance matrices of clusters resulting from the spontaneous clustering. 
Let $ \phi ( x , \mu , \Sigma ) $ be the density function of the normal distribution with mean vector $ \mu $ and covariance matrix $ \Sigma $. Then $ \phi ( x , \hat{ \mu } _ { \gamma k } , \hat{ \Sigma } _ { \gamma k } ) $ is used as a density estimator of mixture component $ f _ k ( x ) $ in (\ref{mixture}). The result of the spontaneous clustering implies the mixture of normal distributions as an estimator of the density function of the population distribution $ g $ in (\ref{mixture}), 
\begin{eqnarray*}
\hat{ g } _ \gamma ( x ) = \sum _ { k = 1 } ^ { K _ \gamma } \hat{ \tau } _ { \gamma k } \phi ( x , \hat{ \mu } _ { \gamma k } , \hat{ \Sigma } _ { \gamma k } ),
\end{eqnarray*}
where $ \hat{ \tau } _ { \gamma k } $ is an estimator of mixing proportion $ \tau _ k $ defined as the proportion of the observations assigned to the $ k $-th cluster.
The AIC based on $ \hat{ g } _ \gamma $ is defined as follows.
$$
\text{AIC} _ \gamma = -2 \sum _ { i = 1 } ^ n \log \hat{ g } _ \gamma ( x _ i ) + 2 \left\{ K _ \gamma \frac{ p ( p + 3 ) }{ 2 } + K _ \gamma -1 \right\} .
$$
The value of $ \gamma $ minimizing $\text{AIC}_\gamma$ is proposed as the optimal selection of $ \gamma $.

\begin{figure}[e]
\begin{center}
\includegraphics[width=80mm]{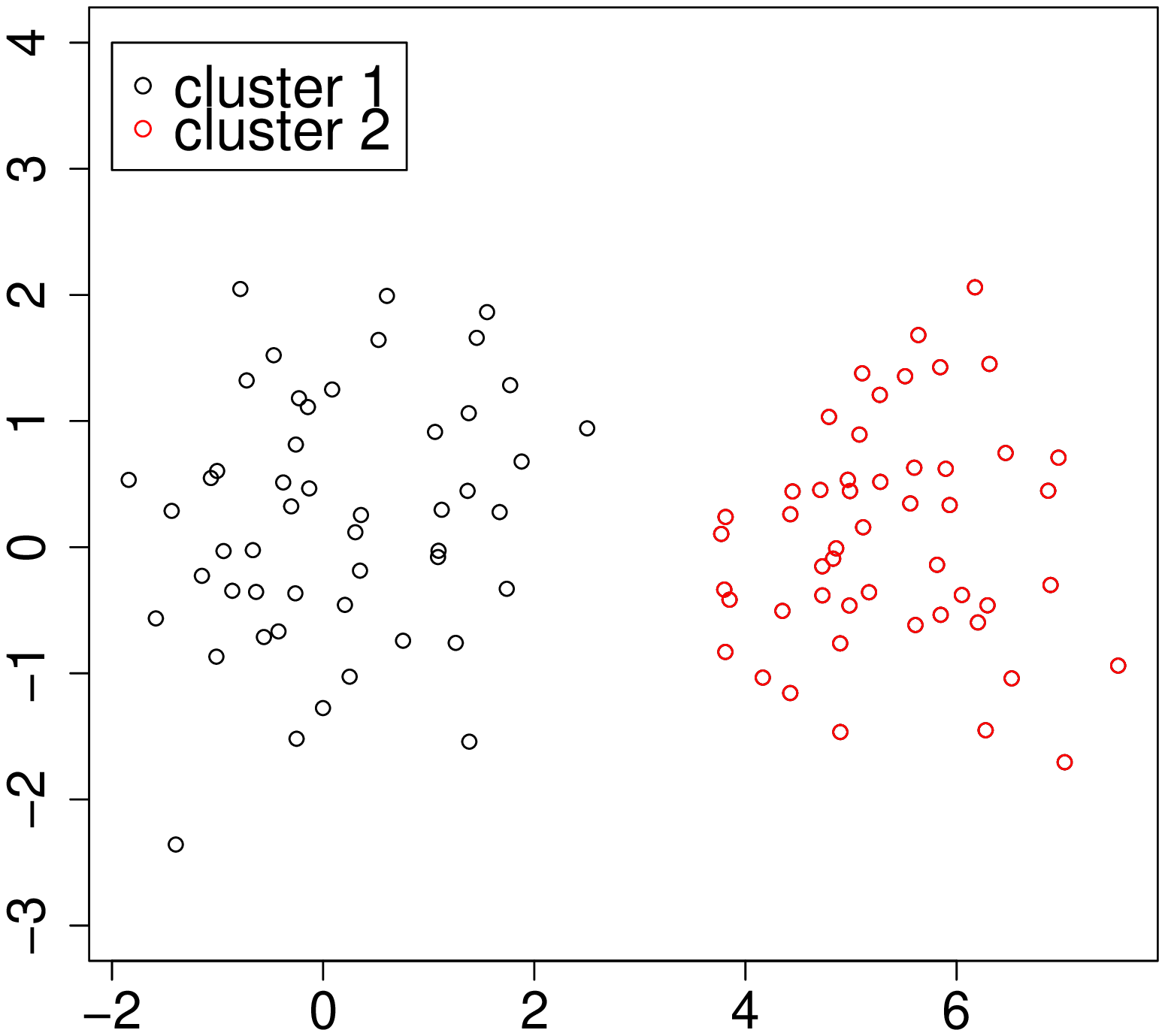}
\end{center}
\caption{Example data generated from the mixture of two normal distributions centered at $ ( 0,0 )^\top $ and $ (5,0)^\top $ with the identity covariance matrix, respectively.} \label{two-clusters}
\end{figure}

\section{Behavior of the $ \gamma $-loss Function}
We provide a justification for the spontaneous clustering by exploring its theoretical aspects. The key fact is that the $ \gamma $-loss function $ L _ \gamma ( \mu ) $ has $ K $ local minimum points if the data set consists of $ K $ cluster groups.

\subsection{Nonconvexity} \label{Difference of two convex functions}
We consider the reason why the $ \gamma $-loss function has local minimum points as illustrated in panel (b) of Figure \ref{robust and clustering}. The optimization problem for a nonconvex function which is expressed as difference of two convex functions has been considered in \cite{label2602} and \cite{label3115}. Effective algorithms such as CCCP and DCA have been developed. Actually, a monotonic transformation of the $ \gamma $-loss function can be expressed as difference of two convex functions, and this expression gives the reason why the $ \gamma $-loss function has local minimum points. Rewrite $ L _ \gamma ( \mu ) $ as  
\begin{eqnarray*}
L _ \gamma ( \mu ) = - \frac{ 1 }{ n }\exp \left[ \log \left\{ \sum _ { i = 1 } ^ n \exp \left( \gamma x _ i ^ \top \mu - \frac{ \gamma }{ 2 } x _ i ^ \top x _ i \right) \right\} - \frac{ \gamma }{ 2 } \mu ^ \top \mu \right] .
\end{eqnarray*}
The local minimum points of $ L _ \gamma ( \mu ) $ are equal to local maximum points of $ \Gamma _ \gamma ( \mu ) = \Gamma _ \gamma ^ { ( 1 ) } ( \mu ) - \Gamma _ \gamma ^ { ( 2 ) } ( \mu )$, where
\begin{eqnarray*}
\Gamma _ \gamma ^ {(1)} ( \mu ) = \log \left\{ \sum _ { i = 1 } ^ n \exp \left( \gamma x _ i ^ \top \mu - \frac{ \gamma }{ 2 } x _ i ^ \top x _ i \right) \right\} , \  \Gamma _ \gamma ^ {(2)} ( \mu ) = \frac{ \gamma }{ 2 } \mu ^ \top \mu . 
\end{eqnarray*}
Then $ \Gamma _ \gamma ^ { ( 2 ) } ( \mu ) $ is obviously a convex function and has a constant Hessian matrix with positive diagonal elements, which means the surface of $ \Gamma _ \gamma ^ { (2) } ( \mu ) $ is curved. 
$ \Gamma _ \gamma ^ { ( 1 ) } ( \mu ) $ is also a convex function because its Hessian matrix is given by 
\begin{eqnarray} \label{curvature}
\frac{ \partial ^ 2 \Gamma _ \gamma ^ { ( 1 ) } ( \mu ) }{  \partial \mu \partial \mu ^ \top  } = \gamma ^ 2 \sum _ { i = 1 } ^ n w ( x _ i , \mu , I ) ( x _ i - \overline{ x } _ { \gamma \mu } ) ( x _ i - \overline{ x } _ { \gamma \mu } ) ^ \top , 
\end{eqnarray}
where $ \overline{ x } _ { \gamma \mu } = \sum _ { i = 1 } ^ n w _ \gamma ( x _ i , \mu , I ) x _ i $, and the Hessian matrix is obviously positive definite. However, the Hessian matrix of $ \Gamma _ \gamma ^ {(1)} ( \mu ) $ varies depending on the data and $ \mu $, and becomes close to the zero matrix in a neighborhood where observations are concentrated. This fact is clear from the form of the Hessian matrix (\ref{curvature}) and means the surface of $ \Gamma _ \gamma ^ {(1)} ( \mu ) $ is almost flat in such a neighborhood. 
Difference between the flat surface and the curved surface causes local maximum points of $ \Gamma _ \gamma ( \mu ) $. Figure \ref{two-graph} illustrates such a phenomenon, where the red, green, and blue lines show $ \Gamma _ \gamma ^ { ( 1 ) } ( \mu ) $, $ \Gamma _ \gamma ^ { ( 2 ) } ( \mu ) $, and $ \Gamma _ \gamma ( \mu ) $, respectively, with dimension $ p = 1 $ and $ \gamma = 3 $. The graphs of $ \Gamma _ \gamma ^ { ( 1 ) } ( \mu ) $ and $ \Gamma _ \gamma ( \mu ) $ are shifted to take 0 at $ \mu = 0 $.  

\begin{figure}[e]
\begin{center}
\includegraphics[width=140mm]{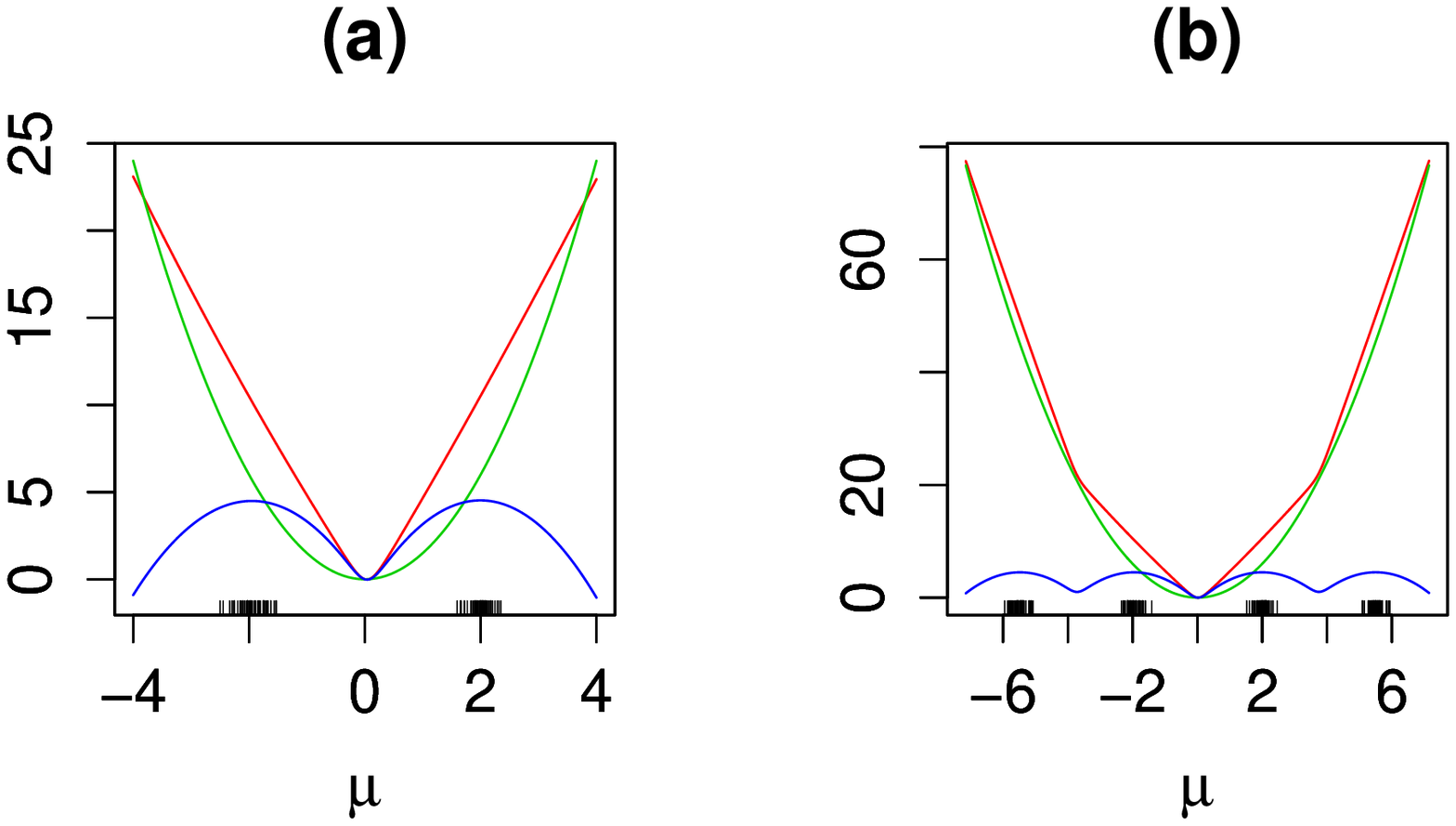}
\end{center}
\caption{Visualization of $  \Gamma _ \gamma ^ { (1) } ( \mu ) $, $  \Gamma _ \gamma ^ { (2) } ( \mu ) $, and $  \Gamma _ \gamma ( \mu ) $. In panel (a) the sample of size 100 is generated from normal mixture $ 0.5 \phi ( x , -2 , 0.04 ) + 0.5 \phi ( x , 2 , 0.04 ) $. In panel (b) the sample of size 200 is generated from normal mixture $ 0.25 \phi ( x , -5.5 , 0.04 ) + 0.25 \phi ( x , -2 , 0.04 ) + 0.25 \phi ( x , 2 , 0.04 ) + 0.25 \phi ( x , 5.5 , 0.04 )  $.} \label{two-graph}
\end{figure}

\subsection{Existence of Local Minimum Points} \label{THE-MOST-IMPORTANT-PROPOSITION}
We consider a condition for the existence of local minimum points of $ L _ \gamma ( \mu ) $. As we discussed in subsection 2.2, the local minimum points of $ L _ \gamma ( \mu ) $ are defined as the centers of clusters, so it is important to know when the $ \gamma $-loss function has local minimum points.

To simplify the argument, we assume that the data set is generated from the mixture of two normal distributions with covariance matrix $ \sigma ^ 2 I   $, 
\begin{eqnarray*}
g ( x ) = \tau _ 1 \phi ( x , \mu _ 1 , \sigma ^ 2 I ) + \tau _ 2 \phi ( x , \mu _ 2 ,  \sigma ^ 2 I ) ,  \ \tau _ 1 + \tau _ 2 = 1 , \ \tau _ k > 0, \ k = 1, 2. 
\end{eqnarray*}
For easy calculation, we consider $ n = \infty $. As $n$ tends to $ \infty $, $ L _ \gamma ( \mu ) $ almost surely converges to the $ \gamma $-cross entropy defined by 
\begin{eqnarray} \label{gammacross}
C _ \gamma ( g , \phi ( \cdot , \mu , I ) ) = - \int g ( x ) \phi ( x , \mu , I ) ^ \gamma dx . 
\end{eqnarray}
See appendix B for the detailed discussion about the $ \gamma $-cross entropy. $ C _ \gamma ( g , \phi ( \cdot , \mu , I ) ) $ becomes
\begin{eqnarray*}
C _ \gamma ( g , \phi ( \cdot , \mu , I ) ) & = & \sum _ { k = 1 , 2 } \tau _ k C _ \gamma ( \phi ( \cdot , \mu _ k ,  \sigma ^ 2 I ) , \phi ( \cdot , \mu , I ) ) \\
&\propto & - \sum _ { k = 1 , 2 } \tau _ k \phi \left( \mu , \mu _ k , \left(  \sigma ^ 2  + \frac{ 1 }{ \gamma } \right) I \right) , 
\end{eqnarray*}
which is nothing but the minus density function of the mixture of two normal distributions with the same covariance matrix $ (  \sigma ^ 2  + 1/\gamma ) I $. Hence the local minimum points of $ C _ \gamma ( g , \phi ( \cdot , \mu , I ) ) $ are equal to the modes of the density function of the normal mixture. Figure \ref{graph-ofcross} shows $- C _ \gamma ( g , \phi ( \cdot , \mu , I ) ) $ with dimension $ p = 2 $, where $ -C _ \gamma ( g , \phi ( \cdot , \mu , I ) ) $ has one or two modes depending on the values of $ \mu _ 1 , \mu _ 2 , \tau _ 1 , \tau _ 2 $, and $ \gamma $. 
For the univariate case, a necessary and sufficient condition that the density function of the mixture of two normal distributions should be bimodal is given in \cite{label6572}. We use a similar technique as in \cite{label6572} to obtain a necessary and sufficient condition for $ C _ \gamma ( g , \phi ( \cdot , \mu , I ) ) $ to have two local minimum points. 
\begin{pro}\label{exist-two-local}
Let $ \nu = ( \mu _ 1 - \mu _ 2 ) / 2 $ and $ d = \| \nu \| ^ 2 - (  \sigma ^ 2  + 1 / \gamma ) $. Then $ C _ \gamma ( g , \phi ( \cdot , \mu , I ) ) $ has two local minimum points if and only if the following three conditions hold: 
\begin{eqnarray}
d  &>& 0, \label{condition1} \\ 
\exp\left( \frac{ 2 \gamma }{ 1 + \gamma \sigma ^2 } \| \nu \| \sqrt{ d } \right) &>&\frac{ \gamma }{ 1 + \gamma \sigma ^ 2 } \left( \| \nu \| + \sqrt{ d } \right) ^ 2  \frac{ \tau _ 1 }{ \tau _ 2 } ,\label{mean-ine1} \\
\exp\left( - \frac{ 2 \gamma  }{ 1 + \gamma \sigma ^ 2 } \| \nu \|  \sqrt{ d } \right) &<& \frac{ \gamma }{ 1 + \gamma \sigma ^2 } \left( \| \nu \| - \sqrt{ d } \right) ^ 2 \frac{ \tau _ 1 }{ \tau _ 2 } \label{mean-ine2}.
\end{eqnarray}
Especially, if $ \tau _ 1 = \tau _ 2 $, then (\ref{mean-ine1}) and (\ref{mean-ine2}) hold for any $ d > 0 $. When the two local minimum points exist, they lie on the segment between $ \mu _ 1 $ and $ \mu _ 2 $. One closer to $ \mu _ 1 $ and the other to $ \mu _ 2 $ are denoted by $ \mu _ 1 ^ * $ and $ \mu _ 2 ^ * $, respectively. Then $ \| \mu _ 1 - \mu _ 1 ^ * \| $ and $ \| \mu _ 2 - \mu _ 2 ^ * \| $ are bounded above by
\begin{eqnarray*}
\| \nu \| - \sqrt{ \| \nu \| ^ 2 - \left( \sigma^2 + \frac{ 1 }{  \gamma } \right) } . 
\end{eqnarray*} 
\end{pro}
By proposition \ref{exist-two-local}, for any $ \sigma ^ 2 $, if $ \mu _ 1 $ and $ \mu _ 2 $ are distinct enough, then there exists $ \gamma $ that guarantees the existence of two local minimum points of $C _ \gamma ( g , \phi (\cdot, \mu,I) ) $, and two clusters are defined at the same instant. In addition, the center of a cluster $ \mu _ k ^ * $ becomes arbitrarily close to $ \mu _ k $ ($k=1,2$), when $ \| \mu _ 1 - \mu _ 2 \| $ becomes large.

\begin{figure}[e]
\begin{center}
\includegraphics[width=65mm]{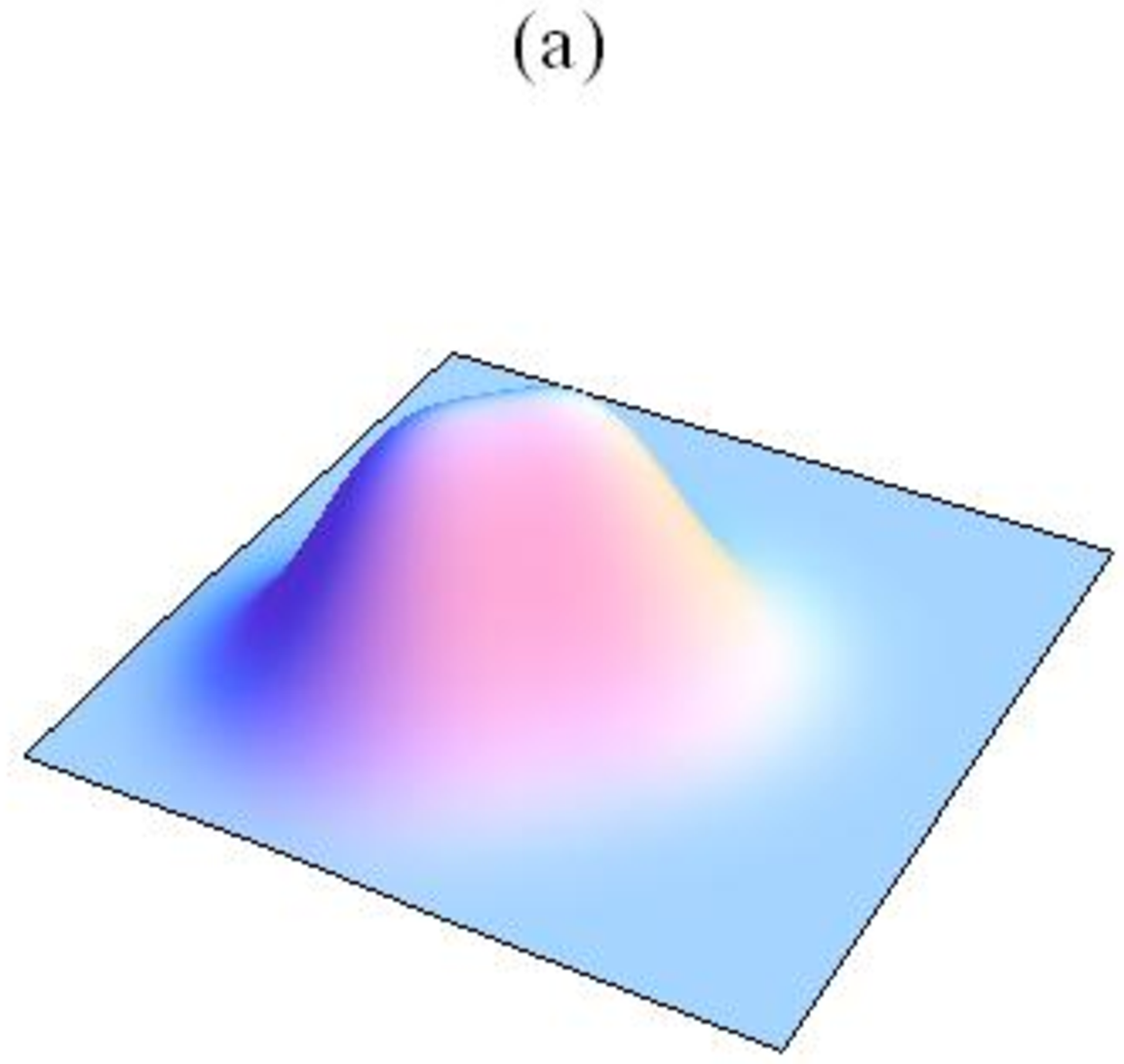}
\includegraphics[width=65mm]{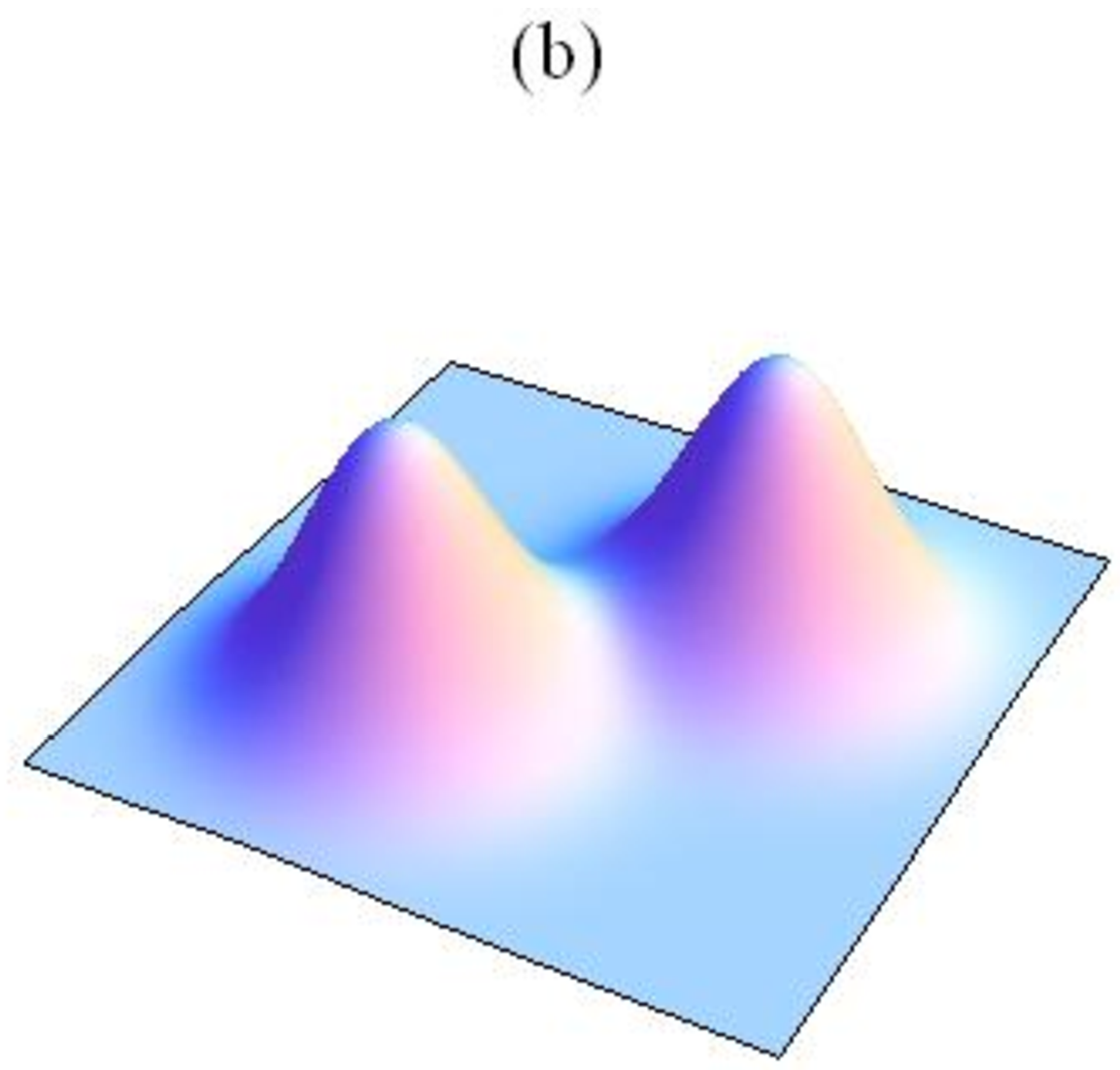}
\end{center}
\caption{Illustration of $ - C _ \gamma ( g , \phi ( \cdot , \mu , I ) ) $. In panel (a) $ \mu _ 1  = ( 0 , 0 ) ^ \top ,  \mu _ 2  = ( 2 , 2 ) ^ \top , \tau _ 1 = \tau _ 2 = 0.5 , \gamma = 1, \sigma^2=1 $. In panel (b) $ \mu _ 1  = ( 0 , 0 ) ^ \top ,  \mu _ 2  = ( 4 , 4 ) ^ \top , \tau _ 1 = \tau _ 2 = 0.5 , \gamma = 1 , \sigma^2=1$.} \label{graph-ofcross}
\end{figure}


\section{Simulation} 
The performance of the spontaneous clustering was investigated through Monte Carlo experiments. 
A comparison of the spontaneous clustering with the $ K $-means algorithm and the model based clustering (MBC) was also implemented.

\subsection{Case of Spherical Clusters}\label{simu1}
We demonstrate the performance of the spontaneous clustering in comparison with the $K$-means algorithm. In this simulation, it is supposed that the covariance matrices of clusters are known to be the identity matrix. 
The value of $ \gamma $ for the spontaneous clustering is determined by the two methods described in subsection \ref{select-gamma}. The number of clusters for the $ K $-means algorithm is determined by two methods described below. The performance of clustering is measured by BHI defined later.

For the $K$-means algorithm, the method by \cite{label6916} and the gap statistic by \cite{label3611} were used to fix the number of clusters. Let $B(k)$ and $W(k)$ be the between- and within-cluster sums of squares with $k$ clusters. 
\cite{label6916} propose to select the number of clusters $ k $ which maximizes $ \text{CH}( k ) $, where $ \text{CH}( k ) $ is defined as 
$$
\text{CH}( k ) = \frac{ B(k) / (k-1) }{W(k)/(n-k)} .
$$
On the other hand,  \cite{label3611} propose to choose the value of $ k $ which maximizes $ \text{Gap}_n(k) = E _ n ^ * ( \log ( W _ k ) ) - \log ( W _ k )$, where $ E _ n ^ * $ denotes expectation under a sample of size $ n $ from the reference distribution. 

The sample of size $ 200 $ is generated from the mixture of five standard normal distributions centered at $ (0,0) ^ \top$, $(3,3) ^ \top$, $ (-3,3) ^ \top, (-3,-3) ^ \top, (3,-3) ^ \top $ with equal mixing proportion. Figure \ref{5-clusters} displays an example sample. 
We simulated 100 runs, and compared clustering results from the spontaneous clustering with those from the $K$-means algorithm. 
Figure \ref{AIC-and-cluster-number.1} shows the value of AIC and the number of clusters resulting from the spontaneous clustering for the sample in Figure \ref{5-clusters}. 
The selected value of $ \gamma $ based on AIC is $ 0.7 $.  

Table \ref{frequencies1} displays the frequency of choosing $K$ clusters for each of the methods for different values of $ K $. 
All methods except the $K$-means algorithm with Gap chose the true number of clusters in almost every simulation run. 
To measure the performance of the clustering, we used Biological Homogeneity Index (BHI) \citep{label7173}, which measures the homogeneity between the cluster $\mathcal C=\{C_1,\ldots, C_K\}$ and the biological category or subtype $\mathcal B=\{B_1,\dots,B_L\}$,
\begin{equation}
{\rm BHI}(\mathcal C,\mathcal B)=\frac1K\sum_{k=1}^K \frac1{n_k(n_k-1)}\sum_{i\not=j,i,j\in C_k}1(B^{(i)}=B^{(j)}),
\end{equation}
where $B^{(i)}\in\mathcal B$ is the subtype for the observation $x_i$ and $n_k$ is the number of the observations in $C_k$. This index is bounded above by 1 meaning the perfect homogeneity between the clusters and the biological categories. 
The mean value of BHI over 100 simulation runs for each method is shown in Table \ref{results1}. 
All methods except the $K$-means algorithm with Gap have good clustering results. 
In every simulation run, if each method detected five clusters for a sample, we calculated the Euclidean distance between the center of a cluster and the mean vector of the corresponding normal component of the normal mixture. 
The mean value of the distance is also shown in Table \ref{results1}, where $\text{DM1}, \ldots , \text{DM5}$ represent the mean value for cluster $1, \ldots , 5$, respectively. 
In this simulation setting, the centers obtained by the spontaneous clustering vary more than those obtained by the $K$-means algorithm. 

To summarize, this simulation example shows that the spontaneous clustering with the range and AIC has almost the same performance as the $K$-means algorithm with CH, and better performance than the $K$-means algorithm with Gap.

\begin{figure}[e]
\begin{center}
\includegraphics[width=140mm]{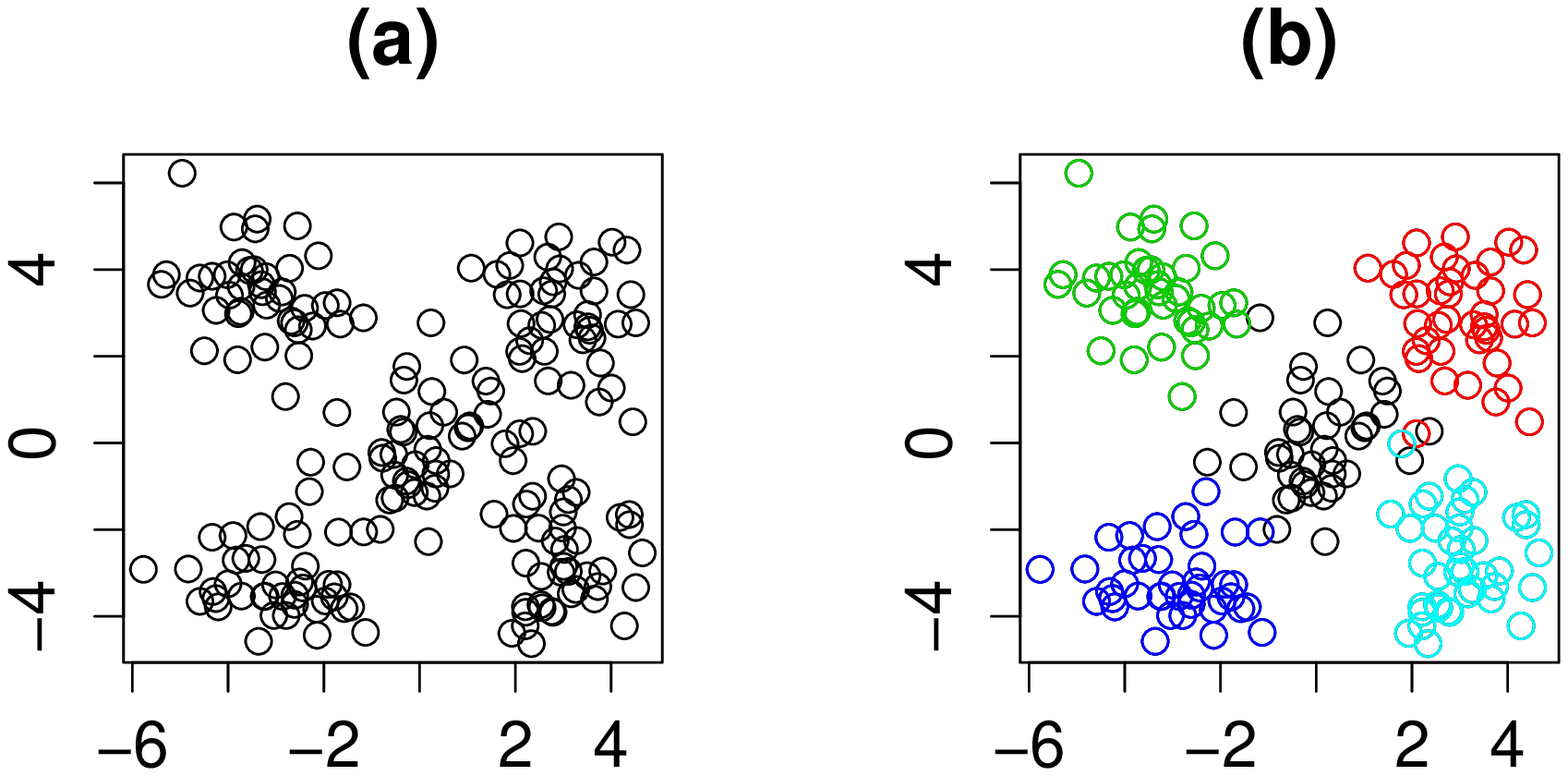}
\end{center}
\caption{(a) Five clusters. (b) Same as (a) but colored according to cluster.} \label{5-clusters} 
\end{figure}
\begin{figure}[e]
\begin{center}
\includegraphics[width=80mm]{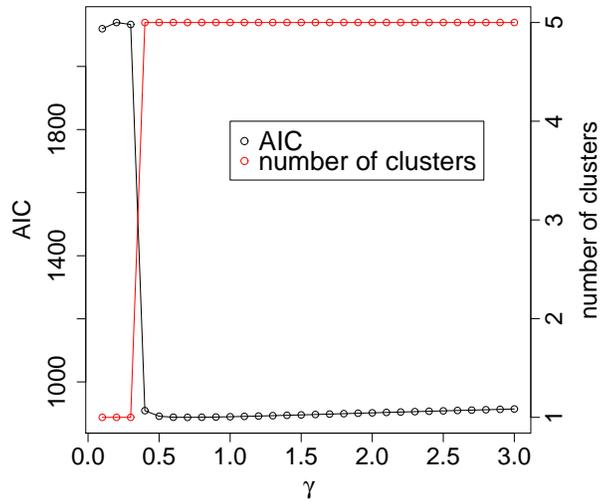}
\end{center}
\caption{Value of AIC and number of clusters.} \label{AIC-and-cluster-number.1}
\end{figure}
\begin{table}[e]
\begin{center}
\caption{Frequencies of Choosing $K$ Clusters.} \label{frequencies1}
\begin{tabular}{crrrrrrrr} 
\hline
$K$ & 1 & 2 & 3 & 4 & 5 \\ 
\hline
Spontaneous clustering with the range & 0 & 0 & 0 & 9 & 91 \\ 
Spontaneous clustering with AIC & 0 & 0 & 0 & 1 & 99 \\ 
$K$-means with CH  & 0 & 0 & 0 & 0 & 100 \\ 
$K$-means with Gap  & 91 & 7 & 0 & 0 & 2 \\ 
\hline
\end{tabular}
\end{center}
\end{table}
\begin{table}[e]
\begin{center}
\caption{Mean Value of BHI and DM1-DM5.} \label{results1} 
\begin{tabular}{crrrrrrrr}
  \hline
& BHI  & DM1 & DM2 & DM3 & DM4 & DM5 \\ 
\hline
Spontaneous clustering with the range & 0.93 & 0.38  & 0.38 & 0.37 & 0.33 & 0.34 \\ 
Spontaneous clustering with AIC & 0.94  & 0.34 & 0.32 & 0.28 & 0.27 & 0.26 \\ 
$K$-means with CH  & 0.95  & 0.25 & 0.23 & 0.21 & 0.21 & 0.21 \\ 
$K$-means with Gap & 0.22 & 0.16 & 0.49 & 0.23 & 0.41 & 0.21  \\ 
\hline
\end{tabular}
\end{center}
\end{table}

\subsection{Case of Ellipsoidal Clusters}\label{Case of Ellipsoidal Heterogeneous Clusters}

We demonstrate the performance of the spontaneous clustering in comparison with the MBC, in which the component density is normal. 
It is supposed that the covariance matrices of clusters are heterogeneous and unknown. 
The value of $ \gamma $ for the spontaneous clustering and the number of clusters for the MBC are determined based on AIC.

The sample of size $ 100 $ is generated from the mixture of two bivariate normal distributions with mean vectors $ (0,0) ^ \top$, $(3,3) ^ \top$, and covariance matrices 
$$
\begin{pmatrix}
1 & 0.5 \\
0.5 & 1 
\end{pmatrix} , \ 
\begin{pmatrix}
2 & -0.5 \\
-0.5 & 2 
\end{pmatrix} .
$$
Figure \ref{2-clusters} displays an example sample, and Figure \ref{AIC-and-cluster-number} shows the value of AIC and the number of clusters resulting from the spontaneous clustering for the sample.
Note that we use two values $ \gamma _ 1 $ and $ \gamma _ 2 $ as power index $ \gamma $. 
$ \gamma _ 1 $ is used for $ L _ \gamma ( \mu ) $ when defining the centers of clusters, and $ \gamma _ 2 $ for $ L _ \gamma ( \mu , \Sigma ) $ when defining the covariance matrices. 
The selected values of $ \gamma_1 $ and $ \gamma_2 $ for the sample in Figure \ref{2-clusters} are $ \gamma_1 = 0.25 $ and $ \gamma _ 2 = 0.7 $.  
We simulated 100 runs, and compared the clustering result from the spontaneous clustering with that from MBC. 

Table \ref{frequencies2} displays the frequency of choosing $K$ clusters for each of the clustering algorithms for different values of $ K $. 
The spontaneous clustering chose the true number of clusters, while the MBC selected large number of clusters 3-10, 39 frequencies. 
The mean value of BHI is shown in Table \ref{result2}. Both clustering algorithms show good performance. In every simulation run, if each clustering method detected two clusters for a sample, two measures were calculated. One is the Euclidean distance between the center of a cluster and the mean vector of the corresponding normal component of the normal mixture. The other is the Frobenius norm of the covariance matrix of a cluster minus that of the corresponding normal component. The mean values of the Euclidean distance and the Frobenius norm are shown in Table \ref{result2}, where DV1 and DV2 represent the mean value of the Frobenius norm for cluster 1 and 2, respectively. In this simulation setting, similar to the simulation result in subsection 4.1, the centers and the covariance matrices obtained by the spontaneous clustering vary more than those obtained by MBC. 

To summarize, this simulation example reveals that the spontaneous clustering with AIC has almost the same performance as MBC with AIC.

\begin{figure}[e]
\begin{center}
\includegraphics[width=140mm]{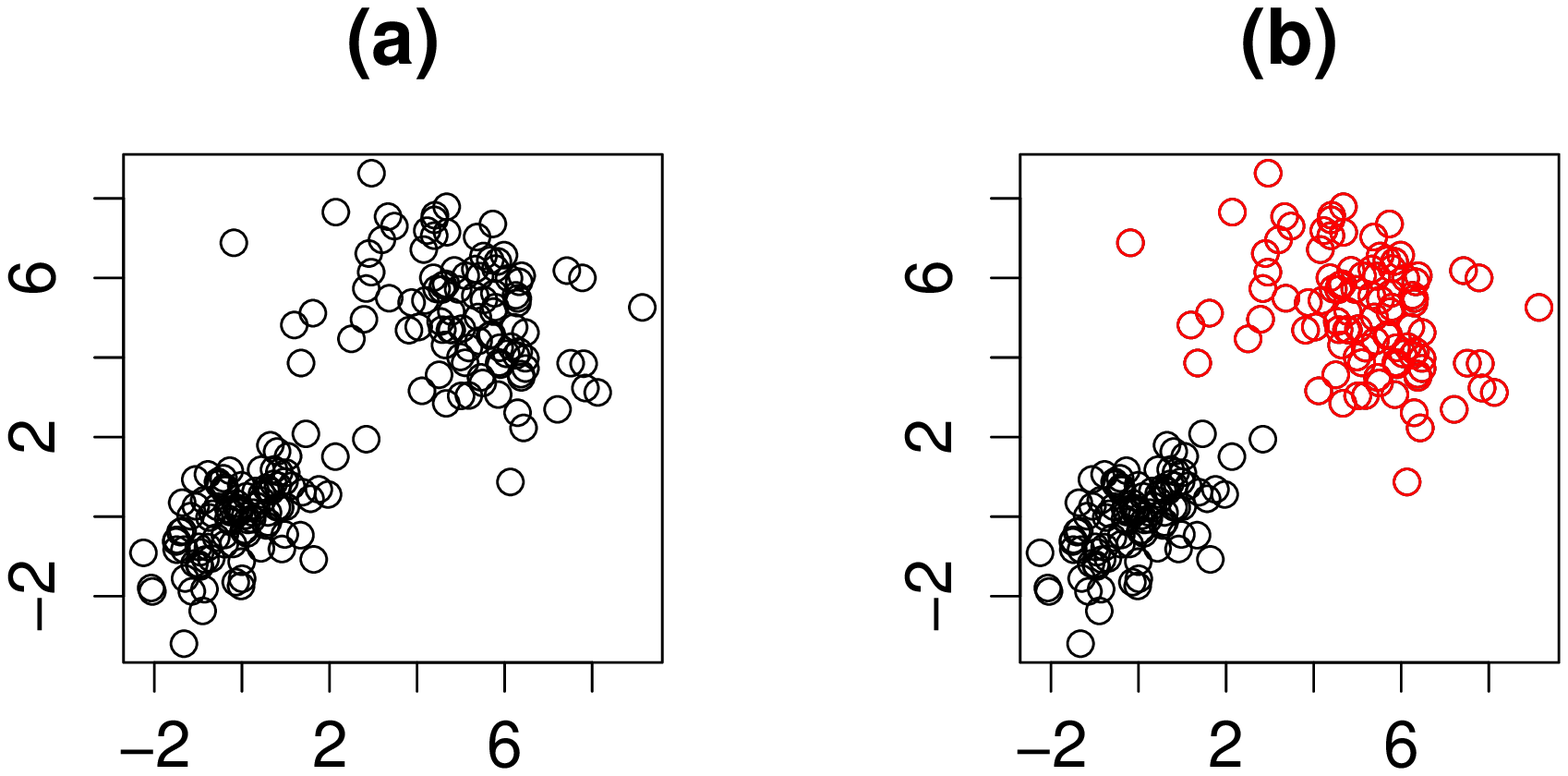}
\end{center}
\caption{(a) Two clusters. (b) Same as (a) but colored according to cluster.} \label{2-clusters}
\end{figure}
\begin{figure}
\begin{center}
\includegraphics[width=140mm]{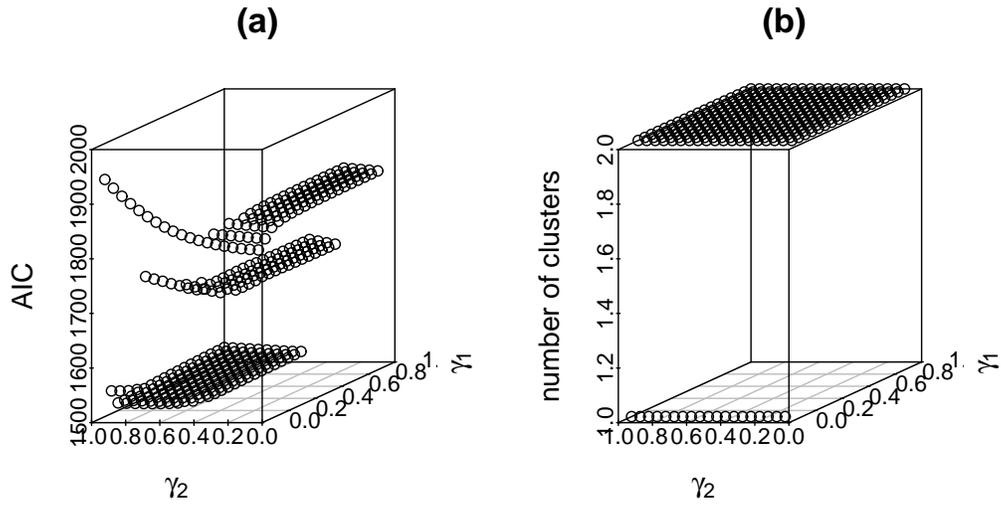}
\end{center}
\caption{(a) Value of AIC. (b) Number of clusters.} \label{AIC-and-cluster-number}
\end{figure}
\begin{table}[e]
\begin{center}
\caption{Frequencies of Choosing $K$ Clusters.} \label{frequencies2}
\begin{tabular}{crrrrrrrrrrrrrr} 
\hline
$ K $ & 1 & 2 & 3 & 4 & 5 & 6 & 7 & 8 & 9 & 10 \\ 
\hline
Spontaneous clustering & 0 & 100 & 0 & 0 & 0 & 0 & 0 & 0 & 0 & 0  \\ 
MBC  & 0 & 61 & 13 & 3 & 4 & 4 & 3 & 4 & 5 & 3 \\
\hline
\end{tabular}
\end{center}
\end{table}
\begin{table}[e]
\begin{center}
\caption{Mean Value of BHI and DM1, DM2, DV1, and DV2.} \label{result2}
\begin{tabular}{crrrrrrrr}
  \hline
&  BHI & DM1 & DM2 & DV1 & DV2 \\ 
\hline
Spontaneous clustering & 1.00 & 0.12 & 0.20 & 0.33 & 0.58   \\ 
MBC & 0.99 & 0.10 & 0.16 & 0.22 & 0.48   \\ 
\hline
\end{tabular}
\end{center}
\end{table}


\section{Data Analysis}
To evaluate the practical performance of the spontaneous clustering, we applied it with the fixed identity covariance matrix to real data as well as the $ K $-means algorithm. 
The data set consists of the chemical composition of 45 specimens of Romano-British pottery, determined by atomic absorption spectrophotometry, for nine oxides \citep{label2088}. 
Figure \ref{pottery} shows the scatterplot matrix of data on Romano-British pottery. 
In addition to the chemical composition of the specimens, the kiln site at which the specimen was found is known. There exist five kiln sites, and they are from three different regions, so that we use the three regions as class labels.  
Our aim is to partition the 45 specimens into clusters corresponding to the three classes by using only information about the chemical composition without knowledge about the class labels. 
The value of $ \gamma $ for the spontaneous clustering is determined by the two methods based on the range of the data and AIC, respectively. The number of clusters for the $ K $-means algorithm is determined by CH and Gap. 

Table \ref{result3} shows the result of the spontaneous clustering. The value of AIC and the number of clusters are shown in panel (a) of Figure \ref{AIC}. With optimal values of $ \gamma $ based on the range and AIC, the spontaneous clustering detects three clusters corresponding to the three regions. In particular, the clustering result by the heuristic choice of $ \gamma $ is the most correct. The scatterplot of $ \text{Al}_2\text{O}_3 $ variable suggests that the number of clusters is two, and the maximum range is obtained from the variable. This is associated with the scenario discussed in the derivation of the heuristic method, in which we assume the number of clusters is two. The values of CH and Gap are shown in panels (b) and (c) of Figure \ref{AIC}. They increase almost monotonically as the number of clusters increases, so CH and Gap do not work well for this data. As a result, we observe the spontaneous clustering based on the range and AIC can detect three clusters properly and partition the 45 specimens into clusters corresponding to the three regions.

\begin{figure}[e]
\begin{center}
\includegraphics[width=140mm]{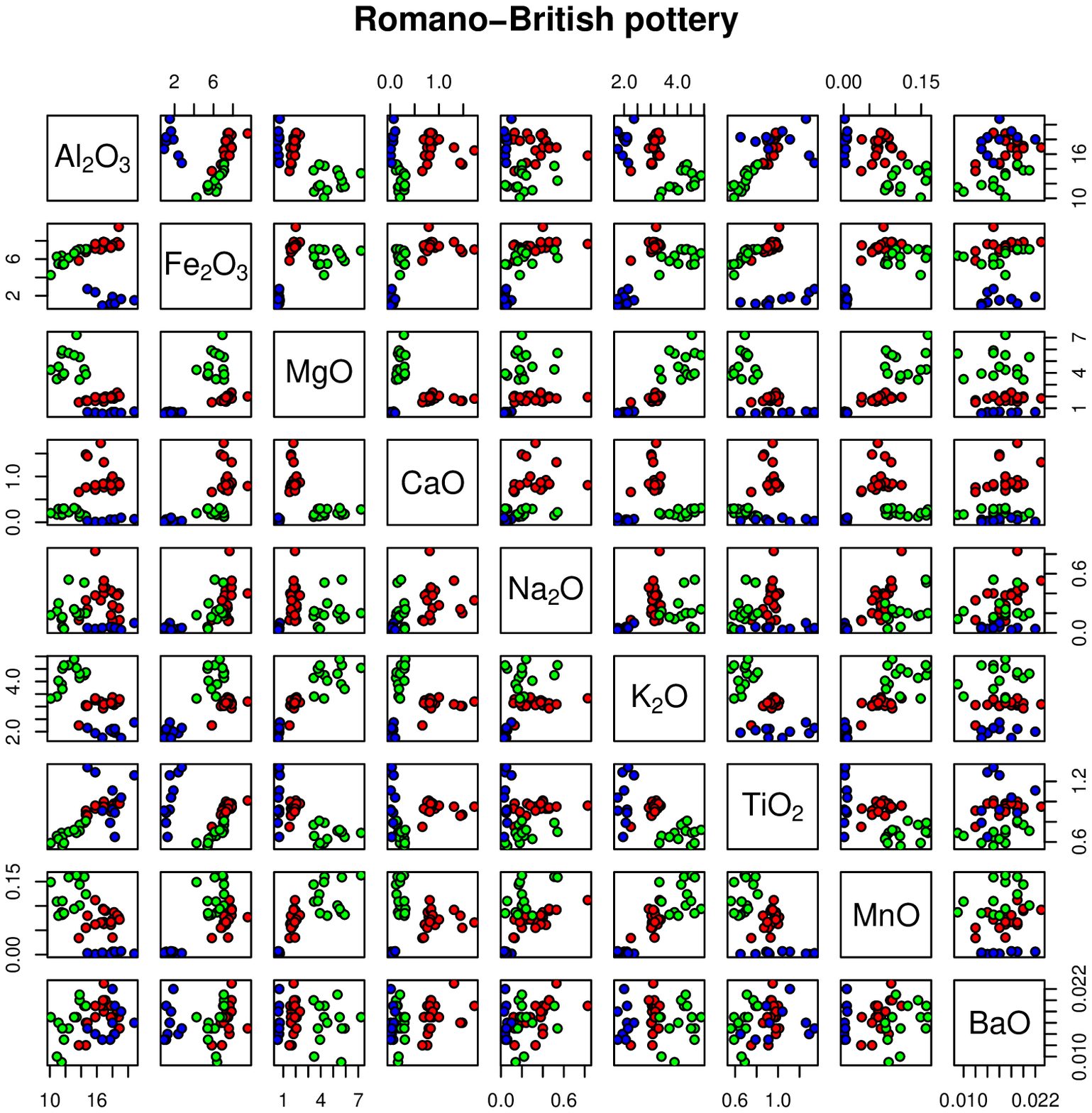}
\end{center}
\caption{Scatterplot matrix of data on Romano-British pottery. The red, blue, and greed circles correspond to the three regions.} \label{pottery}
\end{figure}

\begin{table}[e]
\begin{center}
\caption{Result of the Spontaneous Clustering.} \label{result3}
\begin{tabular}{crrrrrrrr}
\hline
Method & $ \gamma $ & Number of clusters & BHI \\ 
\hline
Range & 0.63 & 3  & 1  \\
AIC & 0.35 & 3 & 0.96 \\
\hline
\end{tabular}
\end{center}
\end{table}

\begin{figure}
\begin{center}
\includegraphics[width=145mm]{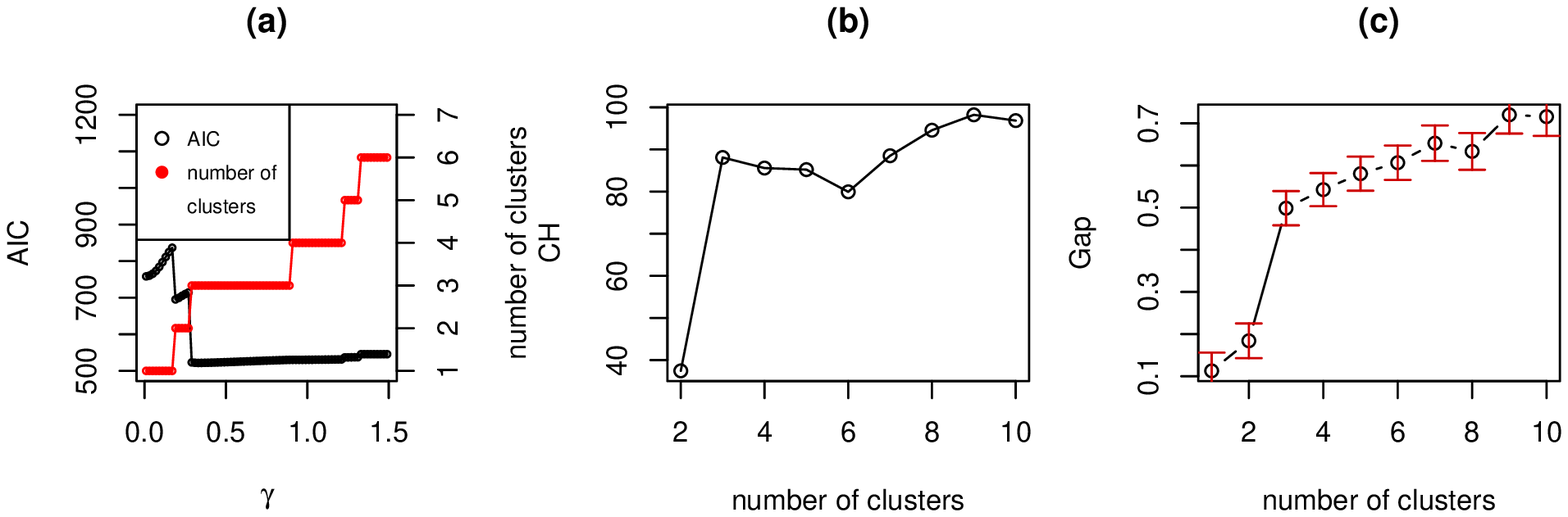}
\end{center}
\caption{(a) AIC and number of clusters. (b) CH. (c) Gap. } \label{AIC}
\end{figure}


\section{Discussion}



We proposed a new clustering algorithm based on the local minimization of the $\gamma$-ross function, which we named the spontaneous clustering. 
In the spontaneous clustering, the local minimum points of the $ \gamma $-loss function are defined as the centers and covariance matrices of clusters. A large majority of statistical methods use the global minimum or maximum point of objective functions and try to avoid local minimum or maximum points. The convexity of the objective functions plays an important role in statistics. For example, support vector machine has a convex loss function, and an efficient algorithm to obtain the global minimum point is considered based on the convexity \citep{label6192}. Although nonconvexity is generally intractable, the spontaneous clustering benefits from the nonconvexity, which makes our method unique and interesting. The idea to use local minimum points of the $ \gamma $-loss function can be applied to other statistical methods. For example, the idea is applied to principal component analysis \citep{label2155} and to estimation of Gaussian copula parameter \citep{label1546}. 

The spontaneous clustering does not require the information about the number of clusters a priori and can find it automatically if the value of power index $ \gamma $ is properly fixed. In contrast, existing methods such as $ K $-means and model based clustering demand the number of clusters. Instead of the number of clusters, the value of $ \gamma $ has to be determined in the spontaneous clustering. Two methods to determine the value of $ \gamma $ are proposed in this paper. One is a heuristic method which depends on the range of the data. Our simulation research shows that it has good performance in many situations, so we can usually use this heuristic method. A more sophisticated choice based on AIC is also proposed although it requires much computational effort}. In the beginning of the research about selection of $ \gamma  $, we considered a cross validation technique, that is one of the common procedures to select the optimal value of a tuning parameter \citep{label9944}. In \cite{label2155} the method using the cross validation is proposed for selection of $ \gamma $. However, the method does not work well for the spontaneous clustering. Hence we employ AIC for selection of $ \gamma $. It is demonstrated that our proposal works well by the simulation study and the real data analysis.




\appendix 

\section{Proof of Proposition \ref{exist-two-local}}
No generality is lost by assuming $ \mu _ 2 = - \mu _ 1 $. 
The gradient of $ C _ \gamma ( g , \phi ( \cdot , \mu , I ) ) $ is given by 
\begin{eqnarray}
\frac{ \partial C _ \gamma ( g , \phi ( \cdot , \mu , I ) ) }{ \partial \mu } &\propto & \tau _ 1 \phi( \mu , \mu _ 1 , (\sigma^2+ 1/\gamma) I ) ( \mu - \mu _ 1 ) \nonumber \\
&& \hspace{1cm} + \tau _ 2 \phi( \mu , -\mu _ 1 , (\sigma^2+ 1/\gamma) I ) ( \mu + \mu _ 1 ) \label{gradient-mu}. 
\end{eqnarray}
From (\ref{gradient-mu}), every local minimum point of $ C _ \gamma ( g , \phi ( \cdot , \mu , I ) ) $ should exist on the segment between $ - \mu _ 1 $ and $ \mu _ 1 $. 
The Hessian matrix of $ C _ \gamma ( g , \phi ( \cdot , \mu , I ) ) $ is given by 
\begin{eqnarray}
\frac{ \partial ^ 2 C _ \gamma ( g , \phi ( \cdot , \mu , I ) ) }{ \partial \mu \partial \mu ^ \top } &\propto& - \tau _ 1 \phi( \mu , \mu _ 1 , (\sigma^2+ 1/\gamma) I ) \frac{ \gamma }{  1 + \sigma^2 \gamma } ( \mu - \mu _ 1 ) ( \mu - \mu _ 1 ) ^ \top  \nonumber \\
&& - \tau _ 2 \phi( \mu , -\mu _ 1 , (\sigma^2+ 1/\gamma) I ) \frac{ \gamma }{  1 + \sigma^2 \gamma } ( \mu + \mu _ 1 ) ( \mu + \mu _ 1 ) ^ \top  \nonumber \\
&& + \tau _ 1 \phi( \mu , \mu _ 1 , (\sigma^2+ 1/\gamma) I ) I  \nonumber \\
&& + \tau _ 2 \phi( \mu , -\mu _ 1 , (\sigma^2+ 1/\gamma) I ) I \label{Hessian-mu} . 
\end{eqnarray}
Let $ \mu ( t ) = t \mu _ 1 $. From (\ref{Hessian-mu}), $ \mu ( t ) $ is a local minimum point of $ C _ \gamma ( g , \phi ( \cdot , \mu , I ) ) $ if and only if $ t $ is a local minimum point of $ C _ \gamma ( g , \phi ( \cdot , \mu( t ) , I ) ) $ with respect to $t$. $ C _ \gamma ( g , \phi ( \cdot , \mu ( t ) , I ) ) $ becomes
\begin{eqnarray*}
C _ \gamma ( g , \phi ( \cdot , \mu ( t ) , I ) ) \propto  - \tau _ 1 \exp( - C ( t - 1 ) ^ 2 ) - \tau _ 2 \exp( - C ( t + 1 ) ^ 2 ), 
\end{eqnarray*}
where $ C $ is equal to $ \| \mu _ 1 \| ^ 2 \gamma / ( 2 ( 1 + \sigma^2\gamma ) ) $. The derivative of $ C _ \gamma ( g , \phi ( \cdot , \mu ( t ) , I ) ) $ is given by 
\begin{eqnarray*}
\frac{ d }{ d t } C _ \gamma ( g , \phi ( \cdot , \mu ( t ) , I ) ) &\propto&  \tau _ 1 \exp( - C ( t - 1 ) ^ 2 ) ( t - 1 ) + \tau _ 2 \exp( - C ( t + 1 ) ^ 2 ) ( t + 1 ). 
\end{eqnarray*}
It is possible to restrict $ -1 < t < 1 $. Then 
\begin{eqnarray}
&& \frac{ d }{ d t } C _ \gamma ( g , \phi ( \cdot , \mu ( t ) , I ) ) > 0 \nonumber \\
& \Longleftrightarrow  & \exp \left( - C ( t + 1 ) ^ 2 + C ( t - 1 ) ^ 2 \right) > \frac{ (1 - t) \tau _ 1 }{ (t + 1) \tau _ 2 } \nonumber \\
& \Longleftrightarrow  & -4 C t + \log( t + 1 ) - \log( 1 - t ) - \log \frac{ \tau _ 1 }{ \tau _ 2 }> 0 \label{equ11}.
\end{eqnarray}
Let $ h( t ) $ be the left hand side of inequality (\ref{equ11}). The derivative of $ h ( t ) $ is given by 
\begin{eqnarray*}
h'(t) = - 4 C + \frac{ 1 }{ t + 1 } + \frac{ 1 }{ 1 - t }, 
\end{eqnarray*}
and
\begin{eqnarray*}
h'(t) > 0 & \Longleftrightarrow  & -4 C ( 1 - t ^ 2 ) + ( 1 - t ) + ( 1 + t ) > 0 \\
& \Longleftrightarrow  & t ^ 2 - \left( 1 - \frac{ 1 }{ 2 C } \right) > 0 . 
\end{eqnarray*}
If $ 1 - 1/ (2C) \leq 0 $, then $ h ' ( t ) \geq 0 $, and $ C _ \gamma ( g , \phi ( \cdot , \mu ( t ) , I ) ) $ has one local minimum point. Hence $ C _ \gamma ( g , \phi ( \cdot , \mu ( t ) , I ) ) $ has two local minimum points if and only if 
\begin{eqnarray*}
1- \frac{ 1 }{ 2 C } > 0 , \ h ( - D ) > 0 , \ h ( D ) < 0,
\end{eqnarray*}
where $ D $ is the positive solution of equation $ h ' ( t ) = 0 $, that is $ D = \sqrt{ 1- 1/(2C) } $. Condition $ 1 - 1 / ( 2 C ) > 0 $ is equivalent to $ \| \mu _ 1 \| ^ 2 - ( \sigma^2 + 1 / \gamma ) > 0 $. Condition $ h ( - D ) > 0  $ is equivalent to 
\begin{eqnarray*}
&&\exp\left( \frac{ 2 \gamma }{ 1 + \sigma^2 \gamma } \| \mu _ 1 \| \sqrt{ \| \mu _ 1 \| ^ 2 - \left( \sigma^2 + \frac{ 1 }{ \gamma } \right) } \right)\\
&& \hspace{1cm} > \frac{ \gamma }{ 1 + \sigma^2 \gamma } \left( \| \mu _ 1 \| + \sqrt{ \| \mu _ 1 \| ^ 2 - \left( \sigma^2 + \frac{ 1 }{ \gamma } \right) } \right) ^ 2  \frac{ \tau _ 1 }{ \tau _ 2 } ,
\end{eqnarray*}
and condition $ h (  D ) < 0  $ is equivalent to 
\begin{eqnarray*}
&& \exp\left( - \frac{ 2 \gamma }{ 1 + \sigma^2 \gamma } \| \mu _ 1 \| \sqrt{ \| \mu _ 1 \| ^ 2 - \left( \sigma^2 + \frac{ 1 }{ \gamma } \right) } \right) \\
&& \hspace{1cm} < \frac{ \gamma }{ 1 + \sigma^2 \gamma } \left( \| \mu _ 1 \| - \sqrt{ \| \mu _ 1 \| ^ 2 - \left( \sigma^2 + \frac{ 1 }{ \gamma } \right) } \right) ^ 2 \frac{ \tau _ 1 }{ \tau _ 2 } .
\end{eqnarray*}
Note that $ \mu _1 ^ * $ is on the line between $ D \mu _ 1 $ and $ \mu _ 1 $. Similarly $ (-\mu_1)^* $ is on the line between $ -\mu_1 $ and $ -D\mu_1 $. Then 
\begin{eqnarray*}
\|  \mu _ 1 ^ * - \mu _ 1 \| \leq ( 1 - D ) \| \mu _ 1 \| = \| \mu _ 1 \| - \sqrt{ \| \mu _ 1 \| ^ 2 - \left( \sigma^2 + \frac{ 1 }{ \gamma } \right) }.
\end{eqnarray*} 

If $ \tau _ 1 = \tau _2 $, then $ h( \pm1 ) = \pm \infty, h ( 0 ) = 0 $. Condition $ 1 - 1/(2C) > 0 $ is equivalent to $ h'( 0 ) < 0 $. Hence two conditions $ h(-D) > 0, h (D) < 0 $ hold whenever condition $ 1- 1/(2C) > 0 $ holds. 
\finish 


\section{$ \gamma $-divergence and $ \gamma $-loss Function}
The aim of this section is to give a general introduction to the $ \gamma $-divergence and the $ \gamma $-loss function. A more detailed discussion can be found in \cite{label3582}. 

\subsection{$ \gamma $-divergence}
Suppose a random sample is generated from a population distribution with density function $ g $. Let $ \{ f ( \cdot , \theta ) \} $ be a family of density functions indexed by parameter $ \theta $. The $ \gamma $-cross entropy between $ g $ and $ f( \cdot , \theta ) $ is defined as 
\begin{eqnarray*}
C _ \gamma ( g , f( \cdot , \theta ) ) = - \kappa _ \gamma ( \theta ) \int g ( x )  f ( x , \theta ) ^ \gamma d x , 
\end{eqnarray*}
with power index $ \gamma >0 $, where $ \kappa _ \gamma ( \theta ) $ is the normalizing constant defined as 
\begin{eqnarray*}
\kappa _ \gamma ( \theta ) = \left( \int f( x , \theta ) ^ { 1 + \gamma } d x \right) ^ { - \frac{ \gamma }{ 1 + \gamma } } .
\end{eqnarray*}
The Boltzmann-Shannon cross entropy between $ g $ and $ f ( \cdot , \theta ) $ is defined by 
$$
- \int g ( x ) \log f ( x , \theta ) d x . 
$$ 
The $ \gamma $-cross entropy and the Boltzmann-Shannon cross entropy have the following relation since $ \kappa _ \gamma ( \theta ) $ converges to 1 if $ \gamma $ tends to $ 0 $.
\begin{eqnarray*}
\lim _ { \gamma \to 0 } \frac{ C _ \gamma ( g , f( \cdot , \theta ) ) + 1 }{ \gamma } &=&  - \int g ( x ) \lim _ { \gamma \to 0 } \left( \frac{ f ( x , \theta ) ^ \gamma - 1 }{ \gamma } \right) d x \\
&=& - \int g ( x ) \log f ( x , \theta ) d x . 
\end{eqnarray*}
Hence the Boltzmann-Shannon cross entropy can be seen as the $ 0 $-cross entropy, and the $ \gamma $-cross entropy can be regarded as an extension of the Boltzmann-Shannon cross entropy. The $ \gamma $-entropy of $ g $ is defined as $ H _ \gamma ( g ) = C _ \gamma ( g , g ) $; the $ \gamma $-divergence between $ g $ and $ f( \cdot , \theta ) $ is defined as 
\begin{eqnarray*}
D _ \gamma ( g , f( \cdot , \theta ) ) = C _ \gamma ( g , f( \cdot , \theta ) ) - H _ \gamma ( g ) .
\end{eqnarray*}
Note that the $ \gamma $-divergence $D _ \gamma ( g , f( \cdot , \theta ) )$ is nonnegative, and $D _ \gamma ( g , f( \cdot , \theta ) )$ is equal to $0$ if and only if $ \theta $ satisfies that $ g ( x ) = f ( x , \theta ) $ almost everywhere $ x $. From these properties, $D _ \gamma ( g , f( \cdot , \theta ) )$ can be seen as a kind of distance between $ g $ and $ f(\cdot, \theta) $ although it does not satisfy the symmetry. When our aim is to find the closest distribution to $ g $ in model $ \{ f ( \cdot, \theta ) \} $ with respect to the $ \gamma $-divergence, we only have to find the global minimum point of $ D _ \gamma ( g , f ( \cdot, \theta ) )  $ with respect to $ \theta $, which is equal to that of $ C _ \gamma ( g , f ( \cdot , \theta ) )  $.

\subsection{$ \gamma $-loss Function} \label{gamma-loss function}
The $ \gamma $-loss function is defined by an estimator of the $ \gamma $-cross entropy. Let $ \{ x _ 1 ,  x _ 2 $, $ \ldots , x _ n \} $ be a random sample generated from a population distribution with density function $ g $ and $ \{ f ( \cdot , \theta ) \} $ be our statistical model. The $ \gamma $-loss function for $ f ( \cdot , \theta ) $ associated with the $ \gamma $-divergence is given by 
\begin{eqnarray*}	
L _ \gamma ( \theta ) &=& - \kappa _ \gamma ( \theta ) \frac{ 1 }{ n } \sum _ { i = 1 } ^ n f ( x _ i , \theta ) ^ \gamma.
\end{eqnarray*}
We extend the definition of the $ \gamma $-cross entropy to any distributions. For any distribution function $ G $, the $ \gamma $-cross entropy between $ G $ and $ f ( \cdot , \theta ) $ is defined as 
\begin{eqnarray*}
C _ \gamma ( G , f ( , \theta ) ) = - \kappa _ \gamma ( \theta ) \int  f ( x , \theta ) ^ \gamma d G ( x ). 
\end{eqnarray*}
Note that $ L _ \gamma ( \theta ) $ equals $ C _ \gamma ( \hat{ G } , f ( \cdot , \theta ) ) $ with empirical distribution function $ \hat{ G } $, so that $ E( L _ \gamma ( \theta ) ) $ $= C _ \gamma ( g , f ( \cdot , \theta ) ) $, and $ L _ \gamma ( \theta ) $ almost surely converges to $ C _ \gamma ( g , f ( \cdot , \theta ) ) $. The $\gamma$-estimator of $ \theta $ is defined by the global minimum point of $ L _ \gamma ( \theta ) $ \citep{label3582}. From the definition of the $ \gamma $-estimator, it satisfies Fisher consistency. If the density function $ g $ belongs to the statistical model $\{ f ( \cdot , \theta ) \}$, then the $ \gamma $-estimator satisfies asymptotic consistency and normality. The $ \gamma $-loss function and the log likelihood function satisfy the following relation 
\begin{eqnarray*}
\lim _ { \gamma \to 0 } \frac{ L _ \gamma ( \theta ) + 1 }{ \gamma } = - \frac{ 1 }{ n } \sum _ { i = 1 } ^ { n } \log f ( x _ i , \theta ) . 
\end{eqnarray*}
Hence the MLE can be regarded as the $ 0 $-estimator and the $ \gamma $-estimator can be seen as an extension of the MLE. 


\bibliographystyle{apalike(modified)}

\begin{thebibliography}{}

\bibitem[An and Tao, 2005]{label3115}
An, L. T.~H. \& Tao, P.~D. (2005).
\newblock The {DC} ({D}ifference of {C}onvex {F}unctions) programming and {DCA}
  revisited with {DC} models of real world nonconvex optimization problems.
\newblock {\em Annals of Operations Research}, 133:23--46.

\bibitem[Bishop, 2006]{label6192}
Bishop, C.~M. (2006).
\newblock {\em Pattern recognition and machine learning}.
\newblock Springer.

\bibitem[Cali{\'n}ski and Harabasz, 1974]{label6916}
Cali{\'n}ski, T. \& Harabasz, J. (1974).
\newblock A dendrite method for cluster analysis.
\newblock {\em Communications in Statistics - Theory and Methods}, 3(1):1--27.

\bibitem[de~Helguero, 1904]{label6572}
de~Helguero, F. (1904).
\newblock Sui massimi delle curve dimorfiche.
\newblock {\em Biometrika}, 3(1):84--98.

\bibitem[Eguchi and Kato, 2010]{label3582}
Eguchi, S. \& Kato, S. (2010).
\newblock Entropy and divergence associated with power function and the
  statistical application.
\newblock {\em Entropy}, 12:262--274.

\bibitem[Fujisawa and Eguchi, 2008]{label6557}
Fujisawa, H. \& Eguchi, S. (2008).
\newblock Robust parameter estimation with a small bias against heavy
  contamination.
\newblock {\em Journal of Multivariate Analysis}, 99(9):2053--2081.

\bibitem[Hastie et~al., 2009]{label9944}
Hastie, T., Tibshirani, R., \& Friedman, J. (2009).
\newblock {\em The elements of statistical learning: Data mining, inference,
  and prediction}.
\newblock Springer, second edition.

\bibitem[Jain et~al., 1999]{label3134}
Jain, A.~K., Murty, M.~N., \& Flynn, P.~J. (1999).
\newblock Data clustering: a review.
\newblock {\em ACM Computing Surveys}, 31(3):264--323.

\bibitem[Jin et~al., 2011]{label6261}
Jin, D., Peng, J., \& Li, B. (2011).
\newblock A new clustering approach on the basis of dynamical neural field.
\newblock {\em Neural Computation}, 23:2032--2057.

\bibitem[Mollah et~al., 2010]{label2155}
Mollah, M. N.~H., Sultana, N., Minami, M., \& Eguchi, S. (2010).
\newblock Robust extraction of local structures by the minimum
  $\beta$-divergence method.
\newblock {\em Neural Networks}, 23(2):226--238.

\bibitem[Notsu et~al., 2012]{label1546}
Notsu, A., Kawasaki, Y., \& Eguchi, S. (2012).
\newblock Detection of heterogeneous structures on the {Gaussian} copula model
  using projective power entropy.
\newblock {\em submitted}.

\bibitem[Tibshirani et~al., 2001]{label3611}
Tibshirani, R., Walther, G., \& Hastie, T. (2001).
\newblock Estimating the number of clusters in a data set via the gap
  statistic.
\newblock {\em Jounal of the Royal Statistical society: Series B},
  63(2):411--423.

\bibitem[Tubb et~al., 1980]{label2088}
Tubb, A., Parker, A.~J., \& Nickless, G. (1980).
\newblock The analysis of {R}omano-{B}ritish pottery by atomic absorption
  spectrophotometry.
\newblock {\em Archaeometry}, 22:153--171.

\bibitem[Wu, 2011]{label7173}
Wu, H. (2011).
\newblock On biological validity indices for soft clustering algorithms for
  gene expression data.
\newblock {\em Computational Statistics and Data Analysis}, 55(5):1969--1979.

\bibitem[Wu et~al., 2011]{label5766}
Wu, J., Zivari-Piran, H., Hunter, J.~D., \& Milton, J.~G. (2011).
\newblock Projective clustering using neural networks with adaptive delay and
  signal transmission loss.
\newblock {\em Neural computation}, 23:1568--1604.

\bibitem[Xu and Wunsch, 2005]{label2910}
Xu, R. \& Wunsch, D. (2005).
\newblock Survey of clustering algorithms.
\newblock {\em IEEE Transactions on Neural Networks}, 16(3):645--678.

\bibitem[Yuille and Rangarajan, 2003]{label2602}
Yuille, A.~L. \& Rangarajan, A. (2003).
\newblock The concave-convex procedure.
\newblock {\em Neural computation}, 15:915--936.

\end{thebibliography}

\end{document}